\documentclass{article}

\PassOptionsToPackage{numbers, sort&compress}{natbib}
\usepackage[final]{neurips_2021_ml4ps}

\usepackage[T1]{fontenc}    % use 8-bit T1 fonts
\usepackage[colorlinks, citecolor=green!80!black, urlcolor=blue, linkcolor=red]{hyperref}
\usepackage{url}            % simple URL typesetting
\usepackage{booktabs}       % professional-quality tables
\usepackage{amsmath,amssymb,amsthm}       % blackboard math symbols
\usepackage{nicefrac}       % compact symbols for 1/2, etc.
\usepackage{microtype}      % microtypography
\usepackage{xcolor}         % colors
\usepackage{graphicx}
\newcommand{\crea}[1]{{\hat #1}^\dagger}
\newcommand{\ann}[1]{\hat #1}

\newcommand{\ket}[1]{\lvert#1\rangle}
\newcommand{\expec}[1]{\langle#1\rangle}

\title{Dispersive qubit readout with machine learning}

\author{%
  Enrico~Rinaldi \\
  University of Michigan, Ann Arbor MI, USA\\
  and Theoretical Quantum Physics Laboratory, RIKEN, Japan\\
  and iTHEMS, RIKEN, Japan\\
  \texttt{erinaldi.work@gmail.com} \\
   \And
   Roberto Di~Candia \\
   Department  of  Communications  and  Networking,
   Aalto  University, Finland\\
   \texttt{rob.dicandia@gmail.com} \\
   \AND
   Simone~Felicetti \\
   Istituto  di  Fotonica  e  Nanotecnologie,   Consiglio  Nazionale  delle  Ricerche  (IFN-CNR), Italy \\
   \texttt{felicetti.simone@gmail.com} \\
   \And
   Fabrizio~Minganti \\
   Institute of Physics, 
   \'Ecole Polytechnique F\'ed\'erale de Lausanne (EPFL), Switzerland \\
   and Theoretical Quantum Physics Laboratory,  RIKEN, Japan \\
   \texttt{fabrizio.minganti@gmail.com} \\
}

\begin{document}

\maketitle

\begin{abstract}
Open quantum systems can undergo dissipative phase transitions, and their critical behavior can be exploited in sensing applications. For example, it can be used to enhance the fidelity of superconducting qubit readout measurements, a central problem toward the creation of reliable quantum hardware. A recently introduced measurement protocol, named ``critical parametric quantum sensing'', uses the parametric (two-photon driven) Kerr
resonator's driven-dissipative phase transition to reach single-qubit detection fidelity of 99.9\% [\href{https://arxiv.org/abs/2107.04503#}{arXiv:2107.04503}].
%Thanks to the diverging susceptibility at the steady state, by tuning the system in the proximity of criticality it is possible to reach extreme precision in quantum measurements.
%For example, 
%It was shown that the readout fidelity of a single qubit detection reaches 99.9\%.
In this work, we improve upon the previous protocol by using machine learning-based classification algorithms to \textit{efficiently and rapidly} extract information from this critical dynamics, which has so far been neglected to focus only on stationary properties.
These classification algorithms are applied to the time series data of weak quantum measurements (homodyne detection) of a circuit-QED implementation of the Kerr resonator coupled to a superconducting qubit.
This demonstrates how machine learning methods enable a faster and more reliable measurement protocol in critical open quantum systems.
\end{abstract}

\section{Introduction}\label{sec:introduction}

A sensing device is a system (weakly) coupled to a second system, the target, characterized by an unknown parameter.
By observing the response of the sensor to the coupling, one can estimate the value of the parameter of the target \cite{Wiseman_BOOK_Quantum}. 
The larger the response of the sensor, the better the estimation of the unknown parameter.
For this reason, quantum systems around criticality have been proposed as sensors, because the diverging susceptibility characterizing second-order phase transition empowers precise parameter estimation.
Such a metrological advantage of quantum phase transitions, however, is limited by a practical consideration: a diverging susceptibility is associated with a critical slowing down, i.e., the emergence of an infinitely-long timescale.
Thus, the price to pay for increased precision is a diverging measurement time \cite{RamsPRX18}. 
Similarly to thermal and quantum phase transitions, an open quantum system can develop second-order Dissipative Phase Transitions (DPTs) \cite{JinPRX16,RotaPRL19} associated with a diverging susceptibility \cite{RotaPRB17}.
Very little has been done in investigating the metrological properties of DPTs \cite{HeugelPRL19,GarbePRL20}.

Two-photon Kerr resonators are remarkably simple open quantum systems that are at the center of intense experimental research in quantum optics and information \cite{LeghtasScience15,Lescanne2020}.
The two-photon Kerr resonator undergoes a second-order DPT \cite{BartoloPRA16,MingantPRA18_Spectral} where the photon number $\langle{\hat{a}^\dagger \hat{a}}\rangle$ has a non-analytical change.
In Ref.~\cite{dicandia2021critical}, the authors introduced a quantum measurement protocol called ``critical parametric quantum sensing'' using the two-photon Kerr resonator. Using this DPT, the state of a qubit (two-level quantum system) coupled to the sensor can be inferred with very high precision.
Indeed, if the qubit is in the state $\ket{\uparrow}$, after a quick transient dynamics, the resonator is almost empty, while if the state is $\ket{\downarrow}$ the resonator contains many photons.
In a first approximation, a homodyne measurement of the photon field allows assessing the state of a qubit with extremely high fidelity (up to 99.9\%).

This protocol, however, relies on a measurement of the long-time (stationary) properties of the Kerr resonator, thus ignoring all the information coming from the short-time dynamics of the system. From a physical point of view, the importance of short-time measurement is clear when considering both non-ideal dispersive measurement and including qubit dissipation and decoherence (both sources of \emph{noise} in current NISQ-era quantum hardware). Thus, in realistic systems, a measurement protocol must be performed within the qubit coherence time. More generally, growing efforts are dedicated to the development of sensing protocols based on dynamical properties of critical quantum systems~~\cite{tsang_quantum_2013,macieszczak_dynamical_2016,Chu2021,gietka2021exponentially,garbe2021critical,fallani2021learning}.

From a theoretical point of view, characterizing the \textit{average} dynamics of an open quantum system requires nontrivial computations of the spectral properties of the generator of the dissipative dynamics \cite{lidar2020lecture}, making it a formidable challenge.
This is even more true when discussing individual trajectories, which describe the outcomes of single experimental realizations.
Indeed, the complexity of the stochastic noise induces nontrivial correlation properties in the dynamics that cannot be captured by the average dynamics of a Lindblad master equation \cite{BartoloEPJST17,RotaNJP18,MunozPRA19}. 
For this reason, an efficient \textit{dynamic} critical parametric quantum sensor cannot rely on simple estimation strategies based on average properties of the system. 
On the other hand, machine learning algorithms are well suited for such a task and are known to work well even in the presence of experimental noise \cite{genois2021quantumtailored}.

%In these scenarios, the state of the qubit changes along the dynamics randomly changes due to the presence of quantum noise, and accordingly the state of the Kerr resonator changes. 
%Thus, long time or steady-state measures become less reliable.

% Indeed, long-time measurements become irrelevant if the qubit has a sizable probability of undergoing a flip along the dynamics.

% (i) Dynamics is more complicated because of fluctuations [I cannot easily estimate the physics] (ii) Even more so because in actual physical experiments non ideal qubit and non-ideal measurement make time a resource.

In this paper, we present a novel machine learning approach to critical parametric quantum sensing that improves the protocol metrological power by taking into account the dynamical aspect of the DPT.
The motivation for this improvement is the following.
Even though the protocol using the steady state reaches high fidelity, it is based on the assumption of a perfect qubit, and as such it requires waiting for a sufficiently long time before the Kerr resonator reaches its steady state.
In actual experimental realizations, however, time is a resource, and the shorter a measurement takes, the less the qubit is prone to errors.
%In other words, an even better measurement instrument can be engineered by reducing the measurement time.
The procedure we propose is a machine learning-based classification of measurement time series, allowing to extract information on the qubit state from the quantum fluctuations of the Kerr resonator dynamics, even at short times.  
% In Section~\ref{sec:resonator} we demonstrate that this protocol is indeed efficient, although it can be extremely challenging to obtain analytical results contrary to the case for the steady-state properties.
%We resort to of this time series, which allow us to obtain fast and reliable determination of the qubit state.

\section{Two-photon Kerr resonator}\label{sec:resonator}

The Hamiltonian of a two-photon driven Kerr-resonator coupled to a qubit is
\begin{equation}
\label{Eq:KerrRes}
\hat{H}_{\rm Kerr} = \omega \crea{a} \ann{a} +\omega_q \hat{\sigma}^{+}\hat{\sigma}^{-} + \frac{\epsilon}{2} ( \hat a^\dagger{}^2 +  \ann{a}^2 ) + \chi \crea{a}{}^2\ann{a}^2 + g \left( \hat{a}^\dagger \hat{\sigma}^- + \hat{a} \hat{\sigma}^+ \right),
\end{equation} 
where $\hat{a}$($\hat{a}^\dagger$) is the bosonic annihilation(creation) operator and $\hat{\sigma}^-$($\hat{\sigma}^+$) is the lowering(raising) spin operator. 
This model can be realized in various photonic platforms, such as circuit-QED implementation,  where a driven resonator is coupled with a superconducting quantum interference device (SQUID) element~\cite{Krantz_2013, Yamamoto_Kerr}. 
%If the resonator is pumped at a frequency $\omega_p\simeq 2\omega_r$, then Eq.~\eqref{Eq:KerrRes} describes effectively the system, by interpreting  
We define the pump-resonator detuning $\omega$, the effective pump-power $\epsilon$, and the SQUID-induced non-linearity $\chi$.

Photons will continuously escape the Kerr resonator, and the number of photons emitted is directly proportional to the number of photons in the Kerr resonator.
This is described by the Lindblad master equation for the system density matrix \cite{Wiseman_BOOK_Quantum}
\begin{equation}\label{Eq:LME}
 \frac{\partial}{\partial t} \hat{\rho}(t) = -i [\hat{H}, \hat{\rho}(t)] + \Gamma[2\hat a \hat{\rho}(t) \hat a^\dag -\{\hat a^\dag\hat a ,\hat{\rho}(t)\}]
\end{equation}
where $\Gamma\geq0$ is the dissipation rate, and it is used as a characteristic timescale.

\begin{figure}[ht]
  \centering
  \begin{tabular}{cc}
      \includegraphics[width=0.45\textwidth]{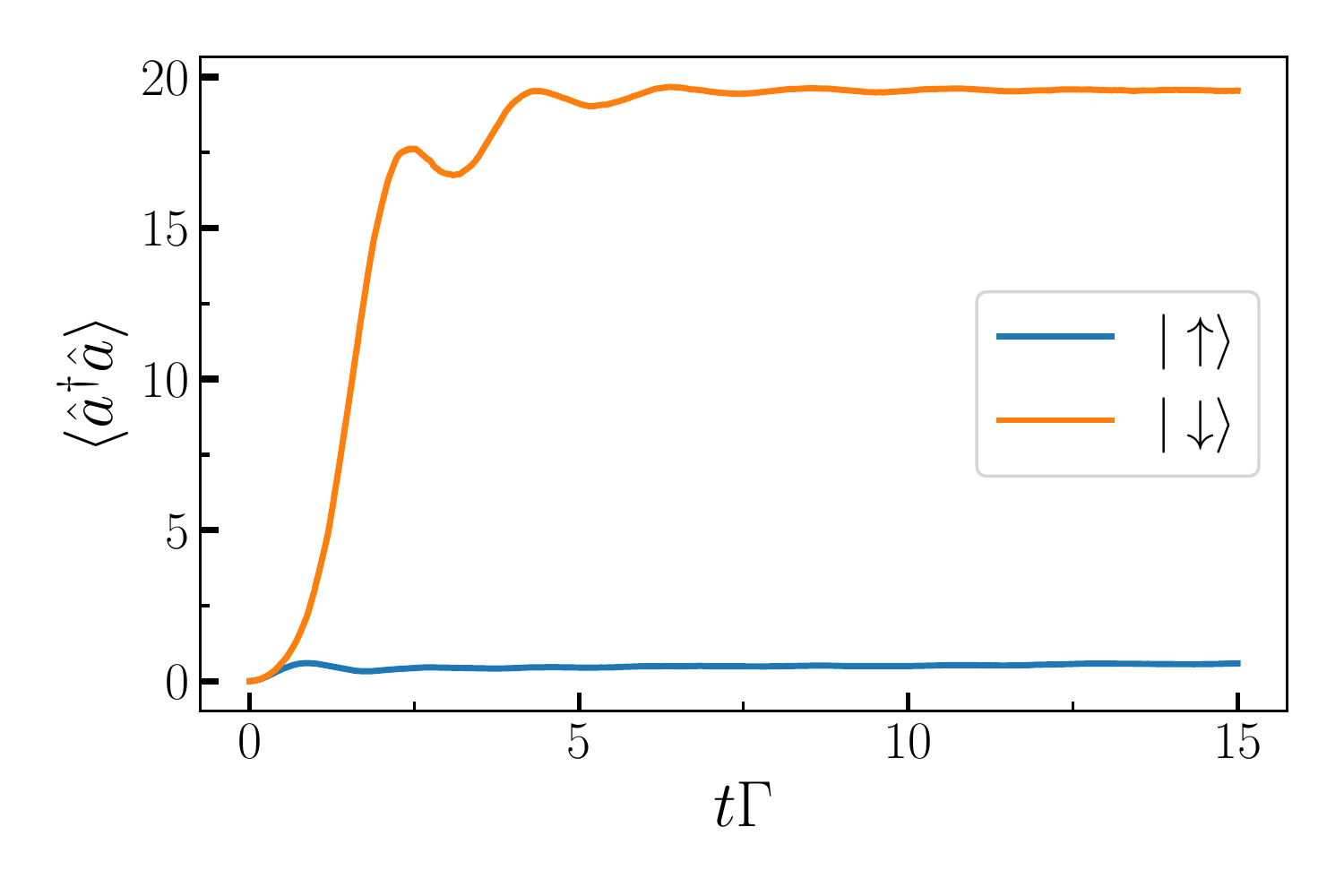} &
      \includegraphics[width=0.45\textwidth]{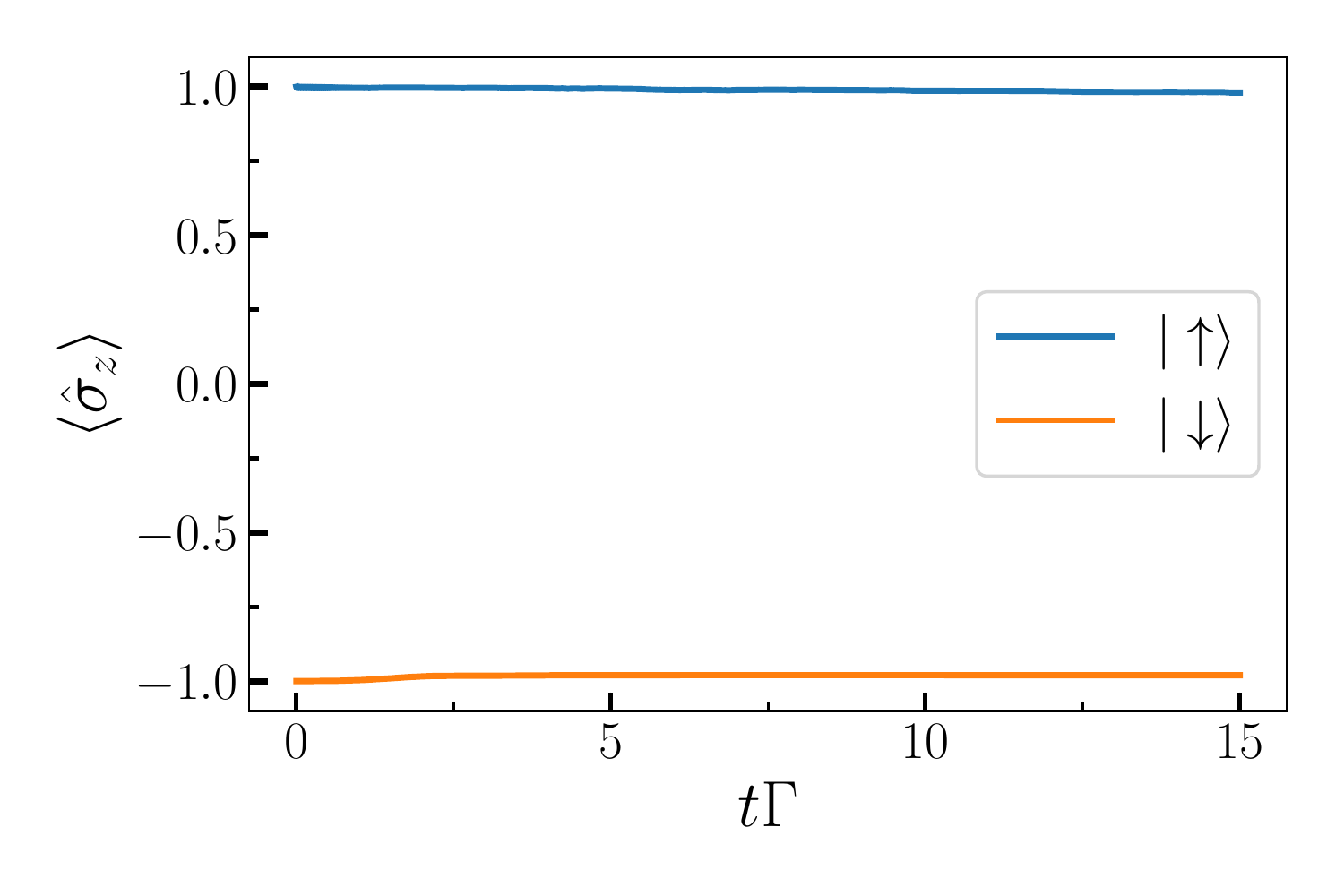}
      \end{tabular}
  \caption{\label{fig:averaged_observables} The number of photons and the qubit state averaged over 1000 quantum trajectories for the two different initial qubit states at $\epsilon=1.67$ and $\delta \omega = 2.3$. The steady-state solution is reached around $t\Gamma \approx 6$.
  }
\end{figure}

Dispersive-readout protocols assume the qubit-resonator coupling to be in the linear dispersive regime, where the qubit-resonator detuning $\Delta=|\omega_q-\omega_r|\gg |\omega_q+\omega_r|/2$. 
In such a regime, the passage of excitation between the qubit and the resonator is suppressed, and the overall effect of the qubit on the resonator is to introduce an effective shift on the frequency of the resonator by $\delta\omega=g^2/\Delta$.
Indeed, if the qubit is $\ket{\uparrow} \,$ ($\ket{\downarrow}$), a photon inside the resonator has energy $\omega + \delta\omega/2 \,$ ($\omega - \delta\omega/2$).
In this dispersive regime, a transition occurs at a critical frequency, such that $\omega  = \sqrt{\epsilon^2 - \Gamma^2}$.\footnote{To be precise,  the DPT occurs in the ``thermodynamic'' limit $\chi \to 0$ \cite{MingantPRA18_Spectral}. Nevertheless, for small $\chi$, a sharp crossover occurs.}
As such, the small change $\delta\omega$ can determine if the system is in the phase with a few photon or in that with many photons, as shown in Fig.~\ref{fig:averaged_observables}, where on the left we show the photon number inside the cavity and on the right we plot the qubit state (see also the discussion in \cite{dicandia2021critical}).
Consequently, by collecting the number of photon emitted, we can determine the state of the qubit.

% The previous discussion took into consideration only the system steady-state properties.
% However, such a steady state is reached after an initial dynamics, which the previous protocol is not considering.
There is, however, another unwanted effect stemming from the coupling to the Kerr resonator.
Indeed, the Kerr resonator induces an effective dissipation on the qubit, introducing additional unwanted errors in the measurement.
Since the probability that a quantum jump induced by the qubit-resonator coupling increases with time, the shorter the required measurement time, the better the outcome.
For example, in Fig.~\ref{fig:flipping_spin} we show that the qubit flips due to the coupling to the cavity. 
Therefore, if we are to consider only the long-time dynamics, in both cases we would wrongly guess the initial state of the qubit.
This is the first reason to consider the Kerr resonator dynamics at short time.

Furthermore, from a metrological point of view, the main theoretical reason for this work is the fact that, contrary to quantum phase transitions, the critical slowing down of DPTs only affects certain observables. 
For instance, at the critical point of the two-photon Kerr resonator, the critical slowing down only affects $\langle{\hat{a}}\rangle$, but leaves unaffected  $\langle{\hat{a}^\dagger \hat{a}} \rangle$. 
This is a common feature of DPTs, and it was demonstrated also for the XYZ model \cite{RotaNJP18}.
Since the protocol proposed in Ref.~\cite{dicandia2021critical} uses photon number to estimate, e.g., the state of a qubit, \emph{it is possible to obtain metrological advantages from the diverging susceptibility of a DPT avoiding the critical slowing down.}
Consequently, the system should display critical fluctuations even at short times, allowing to extract information on the qubit state from the emitted field of the resonator.

\section{Methods}\label{sec:methods}

\begin{figure}[ht]
  \centering
  \begin{tabular}{cc}
      \includegraphics[width=0.45\textwidth]{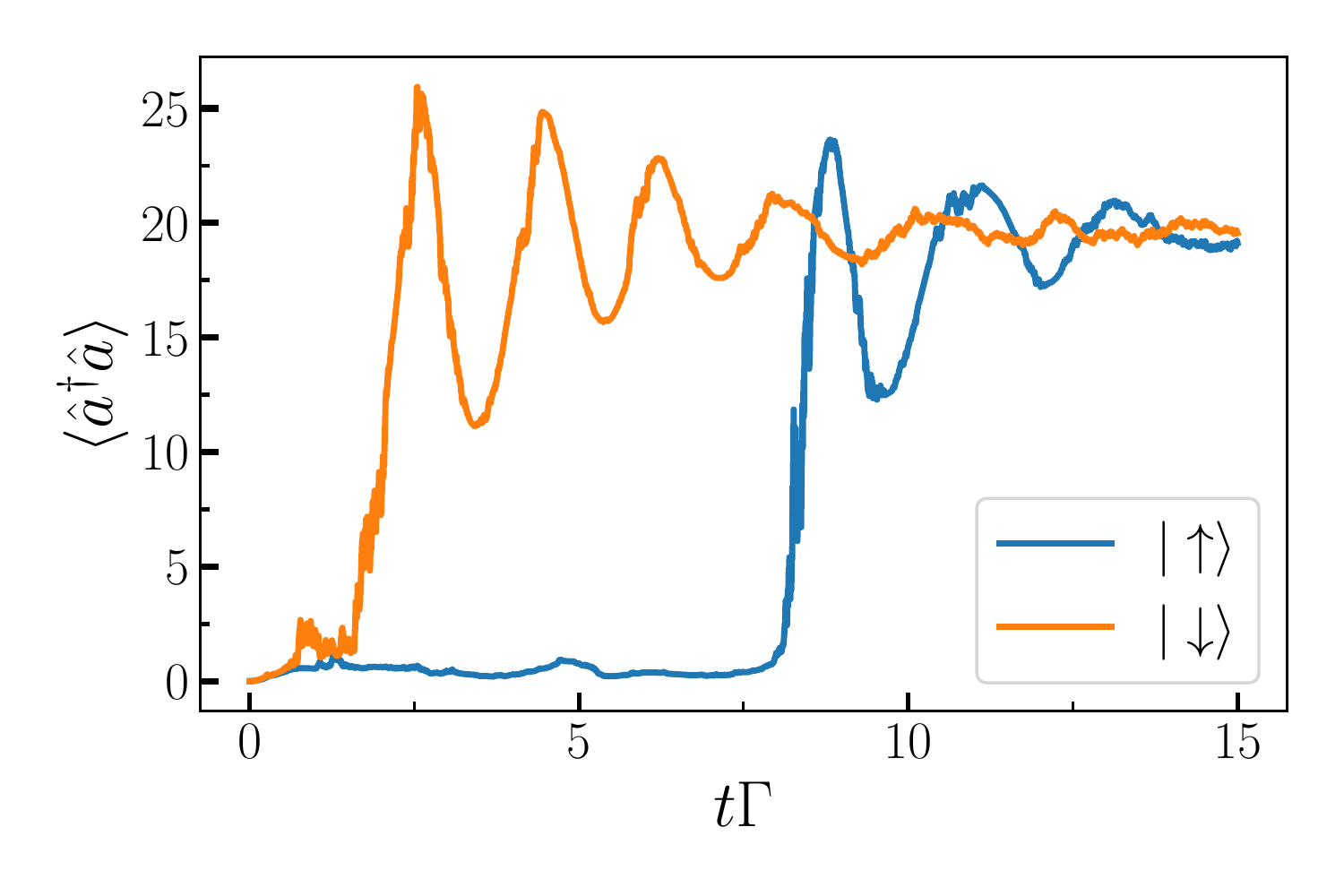} &
      \includegraphics[width=0.45\textwidth]{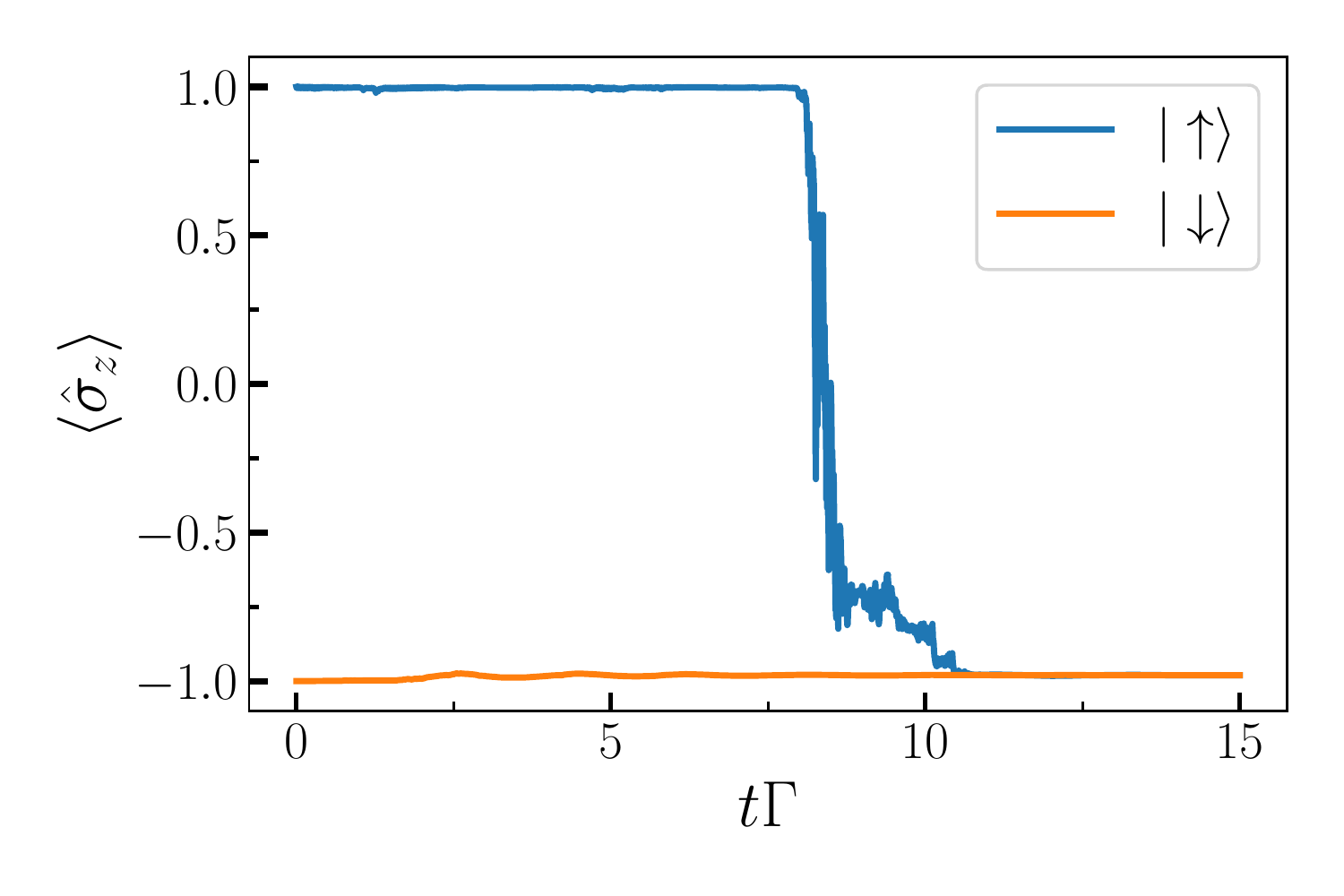}
      \end{tabular}
  \caption{\label{fig:flipping_spin} The number of photons and the qubit state for a single quantum trajectory where one of the qubit states flips (fixed $\epsilon=1.67$ and $\delta \omega = 2.3$). A measurement protocol based on the long-time (stationary) properties would fail to recognize that one qubit was in the $\ket{\uparrow}$ state. Machine learning classifiers can be accurate even with short-time data, when qubits have not had enough time to be altered by the environment.}
\end{figure}

These critical fluctuations can be clearly seen in actual experimental realizations (or their numerical simulations) as shown in Fig.~\ref{fig:flipping_spin} at short time.
By comparing this single-shot dynamics in the left panel, we see important fluctuations of the Kerr-resonator state, far larger than those in the averaged dynamics (for instance, a small peak around $t \Gamma =1$ when the qubit is initialized in $\ket{\downarrow}=\ket{0}$).
This information, however, can be quite challenging to extract from an analytical point of view, because it would require to obtain the Liouvillian superoperator eigenvalues and eigenmatrices (i.e., the spectral decomposition of the generator of the dynamics) \cite{MingantPRA18_Spectral}.
For this reason, we will employ a machine learning algorithm to guess the qubit state exploiting these dynamical critical fluctuations.

Since we are interested in simulating actual experimental measurement outcomes, we resort to homodyne quantum trajectories. 
Instead of the average dynamics of the system in Eq.~\eqref{Eq:LME}, we simulate a set of stochastic Schr\"odinger equations where noise is sampled at each time step \cite{Wiseman_BOOK_Quantum}.
For a fixed set of parameters in $\hat{H}_{\rm Kerr}$ we simulate 1000 quantum trajectories of the system starting from the two different states of the qubit $\ket{0} = \ket{\downarrow}$ and $\ket{1} = \ket{\uparrow}$ using \texttt{QuTiP}~\cite{qutip1,qutip2}.
We save the homodyne quantum evolution for several observables related to the resonator (sensor), such as the number of photons $\expec{\crea{a}\ann{a}}$, and to the qubit (target), such as $\expec{\hat{\sigma}_z}$.

In circuit-QED experiments, the parameter $\Gamma$ is fixed by the manufacturing process, the non-linearity $\chi$ is set by the SQUID characteristics, while the detuning $\Delta$ and the drive intensity $\epsilon$ can be easily tuned in real time.
We choose to save data with a maximal time resolution of $\Delta t = 10^{-3}/\Gamma$ and up to $t\Gamma = 15$, corresponding to state-of-the-art homodyne measurement frequencies.
The other relevant time scales, in units of $\Gamma$, are the total time of measurement $t_f$ and the instrument measurement smoothing scale $\tau$, corresponding to a time interval over which individual values can not be obtained and are averaged over by the measurement instrument.
These two time parameters can be optimized by the protocol to achieve the highest fidelity.

We choose to analyze the quantum trajectories of the observable $\hat{x}$ using two simple machine learning classifiers for time series.
Both classifiers employ a Support Vector Machine algorithm (SVC) with radial basis function kernels (\texttt{rbf}) to discover non-linearity in the feature space.
The first classifier (named \texttt{TAB+SVC(rbf)}) is applied to a feature space identified with all the time points of the trajectory, therefore neglecting the time ordering and time correlations.
On the other hand, the second classifier (named \texttt{RIFE+SVC(rbf)}) acts on a meta-feature space which is constructed from random interval features (RIFE)~\cite{RIFE}, such as the average value or the slope of a random time interval of the trajectory.
To estimate the error on the classification accuracy we utilize repeated cross-validation folds, with 100 repetitions and 5 folds.

Our choice of classifiers is limited and dictated by simplicity.
Similar results should hold irrespective of the classifier algorithm as long as the time series dynamics can be captured.
Ideally, classifiers requiring less data to achieve high accuracy would be preferred because that would improve the experimental setup.
Higher accuracy can be achieved by including careful feature engineering and feature selection steps, which we have left for future investigations.  
% The results of machine learning algorithms should then be compared to baseline accuracy values obtained by simple techniques already in use experimentally.
% Of such baseline techniques we choose a linear classification model based on single features: the average (absolute) value of the measurement along the time series (named \texttt{MEAN+LIN}), or the standard deviation of the measurement along the time series (named \texttt{STD+LIN}).

\section{Results}\label{sec:results}

\begin{figure}[ht]
  \centering
  \begin{tabular}{cc}
      \includegraphics[width=0.45\textwidth]{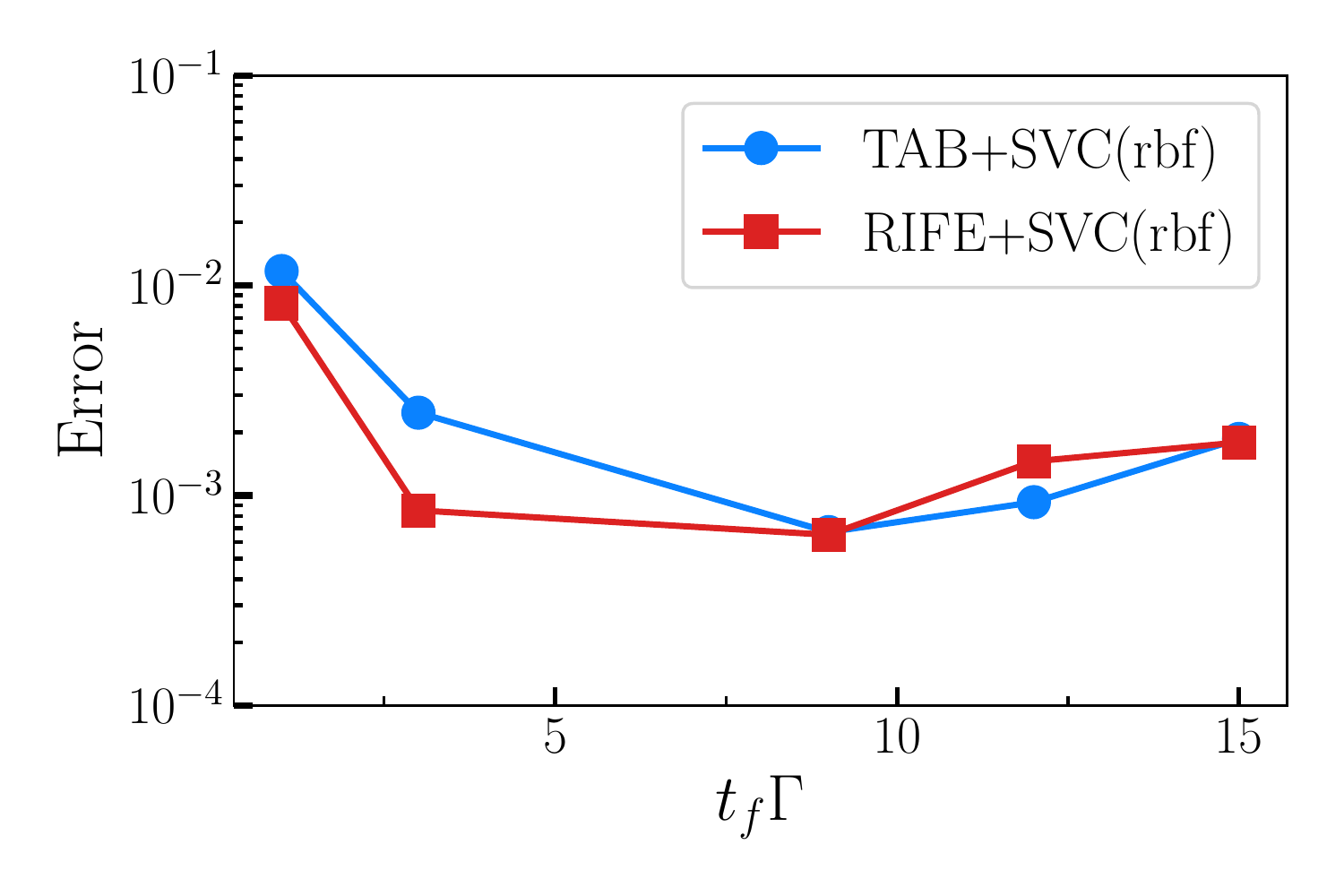} &
      \includegraphics[width=0.45\textwidth]{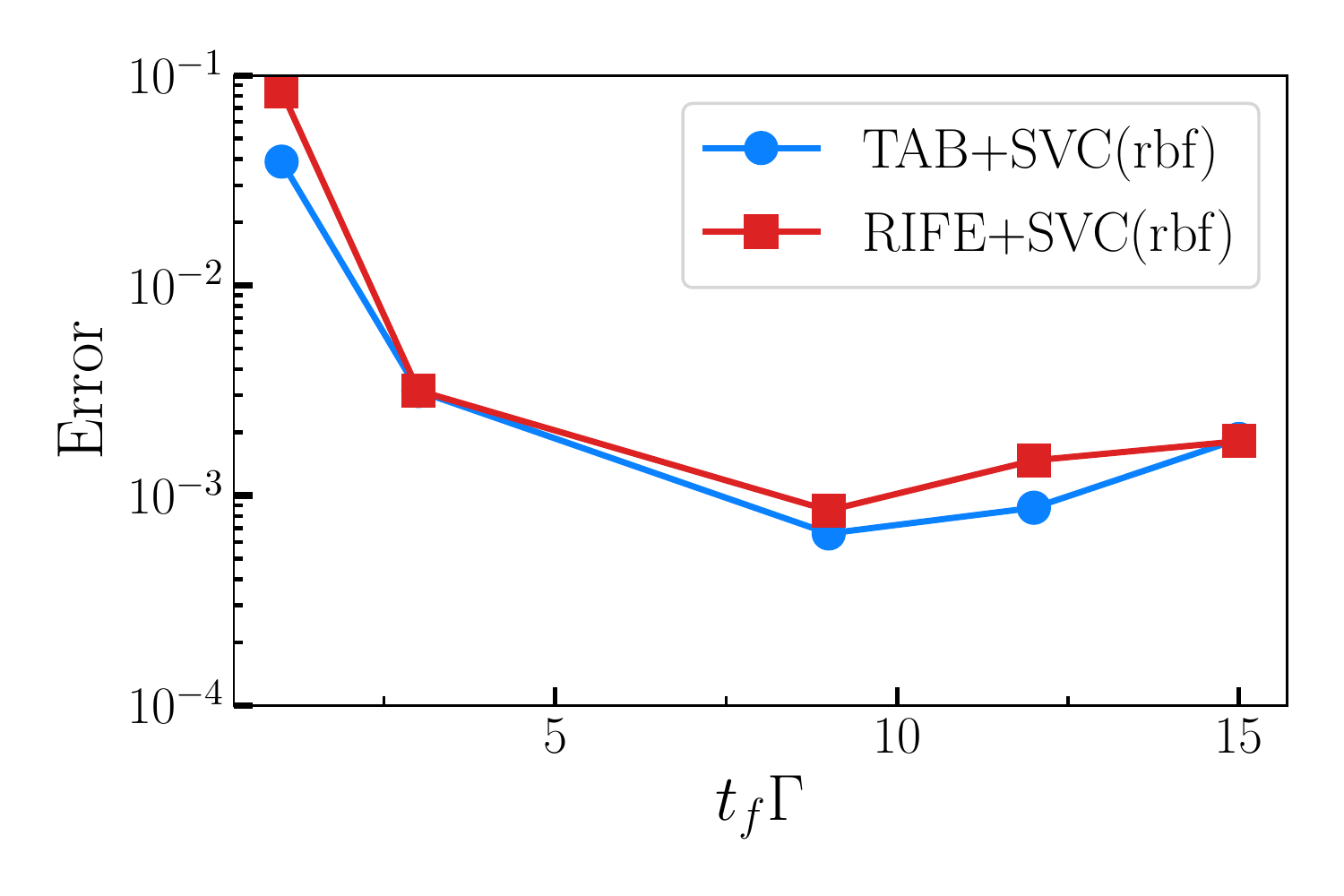}
      \end{tabular}
  \caption{\label{fig:accuracy} The error (1-accuracy) made by different classifiers as a function of the measurement time $t_f\Gamma$ (fixed $\epsilon=1.67$ and $\delta \omega = 2.3$). The left panel is for $\tau\Gamma = 10^{-3}$ and the right panel for $\tau\Gamma = 10^{-1}$.
  }
\end{figure}

We want to tune the Kerr resonator to be the best possible measurement instrument. 
In \citet{dicandia2021critical}, such an optimal point was inferred from analytical considerations on the steady state.
Here instead we consider the dynamical properties.
We scanned the parameter set $\{\epsilon, \, \omega \}$ (which can be easily controlled in standard experimental implementations) and verified the accuracy provided by our choice of classifiers for the state of a qubit initialized either in $\ket{\uparrow}$ or $\ket{\downarrow}$, where the accuracy is the ratio between the number of correct results over the total number of simulations (the variability of the accuracy in our cross-validation procedure is always of the order of 0.002 or less).
This procedure is repeated for different total measurement time $t_f\Gamma$ and for different measurement smoothing scales $\tau\Gamma$: note that this procedure amounts to changing the quality and quantity of samples used in the supervised learning classification task.

In Fig.~\ref{fig:accuracy}, we plot the results obtained at various $t_f\Gamma$.
In the left panel, where $ \tau = 10^{-3}/\Gamma$,  we see that for very short times the algorithm is capable of determining the qubit state in a very short time $ t_f\Gamma  =2$ and very high accuracy (errors $\simeq 1- 9 \times 10^{-4}$).
If we increase $ \tau = 10^{-1}/\Gamma$, we can obtain similar accuracy ($\simeq 1 \times 10^{-3}$), but we need to wait for longer times ($t_f\Gamma =9$).

These results demonstrate that, exploiting critical fluctuations, it is possible to obtain a performing measurement instrument, capable of determining the state of a qubit with extreme high precision and in very short time.
In the future, we plan to further investigate the properties of critical open quantum system on a more mathematical ground, and to test this protocol on actual experimental data, including measurement noise and qubit dissipation effects.

\begin{ack}
We gratefully acknowledge the use of computing time on the Supercomputer HOKUSAI BigWaterfall of the Information Systems Division at RIKEN.\\
E.~R. is supported by Nippon Telegraph and Telephone Corporation (NTT) Research. R.~D. acknowledges support from the Marie Sk{\l}odowska Curie fellowship number 891517 (MSC-IF Green-MIQUEC).

\end{ack}

\bibliographystyle{unsrtnat}
\bibliography{neurips_ml4ps_2021}

\end{document}